\documentclass[12pt,preprint]{aastex}

\shorttitle{Jet Emission, X-ray States and  Hard Tails in GX~17+2}
\shortauthors{Migliari et al.}

\begin{document}


\title{Linking Jet Emission, X-ray States and Hard X-ray Tails in the\\ 
Neutron Star X-ray Binary GX~17+2}


\author{S. Migliari\altaffilmark{1}, J.C.A. Miller-Jones\altaffilmark{2}, R.P. Fender\altaffilmark{3}, J. Homan\altaffilmark{4}, T. Di Salvo\altaffilmark{5}, R.E. Rothschild\altaffilmark{1},
M.P. Rupen\altaffilmark{6}, J.A. Tomsick\altaffilmark{7}, R. Wijnands\altaffilmark{2}, M. van der Klis\altaffilmark{2}} 


\altaffiltext{1}{Center for Astrophysics and Space Sciences, University of California San  Diego, 9500 Gilman Dr., La Jolla, CA 92093-0424}
\altaffiltext{2}{Astronomical Institute `Anton Pannekoek', University of Amsterdam, and
Center for High Energy Astrophysics, Kruislaan 403, 1098 SJ,
Amsterdam, The Netherlands.}
\altaffiltext{3}{ School of Physics and Astronomy, University of Southampton, Hampshire SO17 1BJ, United Kingdom}
\altaffiltext{4}{MIT Center for Space Research, 77 Massachusetts Avenue, Cambridge, MA 02139}
\altaffiltext{5}{Dipartimento di Scienze Fisiche ed Astronomiche, Universita' di Palermo, via Archirafi 36 - 90123 Palermo, Italy}
\altaffiltext{6}{NRAO, Array Operations Center, 1003 Lopezville Road, Socorro, NM 87801, USA}
\altaffiltext{7}{Space Sciences Laboratory/University of California Berkeley, 7 Gauss Way, Berkeley, CA 94720-7450, USA}


\begin{abstract}
We present the results from simultaneous radio (Very Large Array) and
X-ray (Rossi-X-ray Timing Explorer) observations of the Z-type neutron
star X-ray binary GX~17+2.  The aim is to assess the coupling between
X-ray and radio properties throughout its three rapidly variable X-ray
states and during the time-resolved transitions. These observations
allow us, for the first time, to investigate quantitatively the
possible relations between the radio emission and the presence of the
hard X-ray tails and the X-ray state of the source.  The observations
show: 1) a coupling between the radio jet emission and the X-ray state
of the source, i.e. the position in the X-ray hardness-intensity
diagram (HID); 2) a coupling between the presence of a hard X-ray tail
and the position in the HID, qualitatively similar to that found for
the radio emission; 3) an indication for a quantitative positive
correlation between the radio flux density and the X-ray flux in the
hard-tail power law component; 4) evidence for the formation of a
radio jet associated with the Flaring Branch-to-Normal Branch X-ray
state transition; 5) that the radio flux density of the newly-formed
jet stabilizes when also the normal-branch oscillation (NBO) in the
X-ray power spectrum stabilizes its characteristic frequency,
suggesting a possible relation between X-ray variability associated to
the NBO and the jet formation. We discuss our results in the context
of jet models. \end{abstract}


\keywords{radio continuum: general - X-rays: binaries  - stars: individual (GX~17+2) - accretion, accretion disks - ISM: jets and outflows}



\section{Introduction}

In X-ray binaries (XRBs), the studies of relativistic radio jets and
their coupling with X-ray properties received a boost in the last
decade thanks to coordinated multiwavelength observations. Black hole
(BH) systems are surely the best studied among the relativistic jet
XRB sources (see Fender 2006 for a review). However, relativistic jets
are not exclusively associated with BHs. The jet phenomenology
observed in BHs can be found in neutron star (NS) XRBs as well: 1)
highly-accreting NS XRBs can launch {\it transient jets} at
ultra-relativistic velocities (e.g. with bulk Lorentz factors of more
than 15 in Cir~X-1; Fender et al. 2004); 2) low-luminosity NS XRBs can
form a {\it compact jet} of the same kind observed in BH XRBs and
active galactic nuclei (i.e. 4U~0614+091; Migliari et al. 2006). The
necessary ingredients for the formation and launch of relativistic
jets seem not to be related to the nature of the compact object.
Therefore, studies of NS jets and their connections to the accretion
properties have an important impact on our understanding of jet
sources in general.  In BH XRBs, studies of the accretion mode
transitions and their coupling to the jet activity (e.g., Fender,
Belloni \& Gallo 2005) rely on observations of occasional outbursts,
of which there are ~1-2 per year and which usually last a few
months. In the case of NS XRBs and especially in the highly-accreting
class of NSs called `Z-type' (see below), we observe periodic X-ray
state transitions on timescales of a few days.  These state
transitions are thought to be triggered by accretion rate changes.  Z
sources can be considered as the NS `counterparts' of transient
highly-accreting BHs like GRS~1915+105 (see e.g. discussion in
Migliari \& Fender 2005). Studying Z-type NSs, we are therefore able
to assess the evolution of the accretion and jet activity in a very
short observational time and more regularly than in transient BH
sources.

Z-type NS XRBs are a class of seven among the brightest XRBs in our
Galaxy: Sco~X-1, Cyg~X-2, GX~17+2, GX~5-1, GX~340+0, GX~349+2 and
XTE~J1701-462 (eight if we add the `peculiar' Z-source Cir~X-1;
Shirey, Bradt \& Levine 1999). The name of the class derives from the
characteristic `Z' shape they trace in the color-color diagram (CD;
Hasinger \& van der Klis 1989; see van der Klis 2006 for a
review). The three branches which form the Z-shaped CD are called the
Horizontal Branch (HB), Normal Branch (NB) and Flaring Branch (FB),
and define three distinct states of the systems, each with its own
specific X-ray spectral and timing properties (see e.g. Homan et
al. 2002 for the specific case of GX~17+2).  The relations between
spectral and timing properties and the position on the CD suggest the
accretion rate as the major physical parameter behind the position
along the Z track (e.g. van der Klis 2006 for a review). Homan et
al. (2007) showed that the complex variability behaviour of
XTE~J1701-462 could be explained if the changes along the track are
governed, not simply by the mass accretion rate through the disk, but
by the ratio between this quantity and its time-averaged variations
(see also van der Klis 2000).  Z sources show luminosities
persistently near or above the Eddington limit and are very bright and
rapidly variable both in X-rays and in the radio band.  The Z-type NSs
change X-ray states on timescales of hours to days, so they are always
at the `edge' of state transitions.  They are therefore very good
laboratories to study the connection between X-ray properties, state
changes and radio behaviour in X-ray binary systems.  Z-type NSs show
optically thick radio emission (i.e., $\alpha\sim 0$, where
$F_{\nu}\propto
\nu^{\alpha}$ and $F_{\nu}$ is the flux density at the frequency
$\nu$) and frequent optically thin radio flares ($\alpha<0$). The
optically thick emission is usually interpreted as radiation from a
continuously replenished compact jet (see Fender 2006 for a review),
while the optically thin radio flares are possible signatures of fast
ejected plasmons, already observed as extended lobes in Sco~X-1
(Fomalont et al. 2001a,b) and Cir~X-1 (Fender et al. 1999; Fender et
al. 2004).

Looking in detail at the radio behaviour of Z sources as a function of
their X-ray properties, Penninx et al. (1988) first found in GX~17+2
that the radio emission varied as a function of the position in the
X-ray CD, decreasing with increasing inferred mass accretion rate from
the HB (strongest radio emission) to the FB (weakest radio
emission). A behaviour consistent with GX~17+2 has been found also in
Cyg~X-2 (Hjellming et al 1990a) and Sco~X-1 (Hjellming et al 1990b),
the exception seems to be GX~5-1 (Tan et al. 1992; but see discussion
in Migliari \& Fender 2006). Based on the results of these previous
simultaneous radio/X-ray observations and - mostly - on observations
of Sco X-1 (Fomalont et al. 2001b; Bradshaw et al. 2003), a possible
coherent phenomenological picture of Z sources, coupling X-rays
($<20$~keV) and radio properties, has been drawn (Migliari \& Fender
2006). In our sketch, 1) the compact jet is mostly responsible for the
radio emission in the HB and partially in the NB; 2) X-ray state
transitions are coupled with transitions also in the jet emission,
specifically transient radio optically thin flares appear to occur at
the HB-to-NB transition (see Fig.~6 in Migliari \& Fender 2006).

Non-thermal hard tails in the X-ray energy spectra, dominating above
$\sim30$~keV, have been observed in almost all the known Z sources:
GX~5-1 (Asai et al. 1994), GX~17+2 (Di Salvo et al. 2000), Sco X-1
(D'Amico et al. 2001a), GX~349+2 (Di Salvo et al. 2001; see also
D'Amico et al. 2001b), Cir X-1 (Iaria et al. 2001) and Cyg~X-2 (Di
Salvo et al. 2002; see also D'Amico et al. 2001b). These hard X-ray
tails can be fitted with a power law with photon index ranging between
1.6 and 3.3, and can contribute up to 10\% of the bolometric X-ray
luminosity. Although the details of the individual Z sources are more
complicated, we might say that, as a general trend, the hard X-ray
tail in the spectrum seems to be related to the position in the CD:
the hard X-ray component is strongest in the HB and becomes weaker
towards the FB. In previous RXTE observations of GX~17+2, for example,
the spectrum showed a hard power-law tail with a photon index of
$\sim2.7$ which is strongest in the HB and weakens as the source moves
towards the NB, disappearing in the NB (Di~Salvo et
al. 2000). However, a clear counter example seems to be Sco~X-1, where
no obvious relation between the hard tail X-ray flux and the position
on the CD has been observed (D`Amico et al. 2001a).  The physical
origin of this non-thermal component is still an area of
controversy. In particular, two main possibilities are under debate,
both suggesting that the physical site of the emission is the central
core region, close to the compact object: inverse Compton from a
non-thermal electron population in a `corona' (e.g. Poutanen \& Coppi
1998) or in the base of a jet (see e.g. Markoff, Nowak \& Wilms 2005
for a discussion).  Another possible explanation, the bulk motion
Comptonization (BMC) which was first explored by Titarchuck,
Mastichiadis \& Kylafis (1996) for BHs, has been proposed also for NS
systems (e.g. Titarchuk
\& Zannias 1998). In the specific case of GX~17+2, for example,
Farinelli et al. (2007) suggested that the spectrum derived from the
BMC of soft photons by energetic electrons flowing on to the NS, can
produce an X-ray hard tail consistent with the observations.

In this work we present a study of simultaneous radio and X-ray
observations of the Z source GX~17+2, in order to assess the coupling
between X-ray and radio properties throughout its three X-ray states
(\S~3.1) and during the transitions between states
(\S~3.2). Furthermore, these observations allow us, for the first
time, to investigate more quantitatively the possible relations
between the radio emission and the presence of the hard X-ray tails
(\S~3.3), observed to be associated with the position of the source in
the CD.

\section{Observations}

We have observed the Z-type NS XRB GX~17+2 simultaneously in X-rays
with the Rossi X-ray Timing Explorer (RXTE) and in the radio band with
the Very Large Array (VLA), covering all its three X-ray branches.

\subsection{RXTE observations and data analysis}

RXTE observed GX~17+2 starting on 2002 November 04, for a total of
$\sim35$ hr over about 11 days. The 3-20~keV light curve of the
observations is shown in Fig.~1, middle panels. We used the
Proportional Counter Array (PCA) {\tt Standard2} data of the
proportional counter unit 2 (working in all the observations) to
produce the CD and the HID of all the RXTE observations of
GX~17+2. The soft color and the hard color are defined as the count
rate ratio (4.6--7.1)~keV/(2.9--4.6)~keV and
(10.5--19.6)~keV/(7.1--10.5)~keV, respectively.  The HID in Fig.~2
clearly shows the three distinct X-ray branches of the source.

For the spectral analysis, we have extracted X-ray energy spectra
using PCA {\tt Standard2} and HEXTE {\tt Standard Mode} data. For the
PCA data, we subtracted the background estimated using {\tt pcabckest}
v3.0, produced the detector response matrix with {\tt pcarsp} v10.1,
and analysed the energy spectrum in the range 3--25~keV. A systematic
error of 0.5\% was added to account for uncertainties in the
calibration.  For the HEXTE data, we corrected for deadtime,
subtracted the background, extracted the response matrix using FTOOLS
v.6.1.2, and analysed the spectra between 20 and $100$~keV (there is
no significant detection above this energy in any of the
observations).  We show the log of the RXTE observations in Table~1.
The 3-100 keV spectra are well fitted (see also Di Salvo et al. 2000)
using a black-body, a thermal Comptonization model (CompTT), a
Gaussian emission line in the range 6.4--6.7 keV and an edge around 9
keV. An additional power law to account for an excess, of non-thermal
origin, in the higher energy range above 30 keV is also necessary in
three observations (see \S~\ref{hardtail} for a discussion). For a
detailed X-ray spectral analysis of GX~17+2 with other models see
e.g. Farinelli et al. (2005). The best-fit parameters of each of the
3-100 keV spectra as a function of their position on the HID are shown
in Table~2. The values are consistent with what was found in Di Salvo
et al. (2000).

For the temporal analysis, we have used {\it event} and {\it binned}
data with a time resolution of 125 $\mu$s for the production of the
power density spectra. We used time bins such that the Nyquist
frequency is 4096~Hz. For each observation we created power spectra
from segments of 64s length using Fast Fourier Transform techniques
(van der Klis 1989 and references therein), but no background
subtraction was performed. No deadtime corrections were done before
creating the power spectra. We averaged the Leahy-normalised power
spectra (Leahy et al. 1983) and subtracted the predicted Poisson noise
spectrum applying the method of Zhang et al. (1995), shifted in power
to match the spectrum between 3000 and 4000~Hz.  We converted the
normalisation of the power spectra to squared fractional rms
(e.g. Belloni \& Hasinger 1990; see van der Klis 1995).

\subsection{VLA observations and data analysis}
\label{sec:radio_data}

We observed GX\,17+2 with the VLA at three different epochs on 2002
November 4, 9, and 15, simultaneous with the {\it RXTE} observations.
The VLA was in its C configuration at the time.  The NRAO project ID
was AR\,495.  Observing frequencies were 1.425, 4.86, 8.46, 14.94,
22.46 and 43.34\,GHz.  Observations were carried out in standard
continuum mode at each frequency, with a 50-MHz bandwidth in each of
two IF pairs.  The primary calibrator was 3C\,286, used to set the
flux scale according to the coefficients derived at the VLA in 1999 as
implemented in the 31Dec05 version of the National Radio Astronomy
Observatory's (NRAO) Astronomical Image Processing System
(AIPS).  The secondary calibrators were J1834--126 ($4.7^{\circ}$
from the target) at 1.46\,GHz and J1832--105 ($5.3^{\circ}$ from the
target) at all higher frequencies.  Data calibration and imaging were
performed using standard procedures within AIPS.  At frequencies
below 15\,GHz, there were background sources present in the field,
most notably an AGN 88\,arcmin to the southwest, which had to be
properly deconvolved from the image.  The data were subjected to a
single round of phase-only self-calibration before making separate
images of the field in the time intervals corresponding to the
different X-ray states.  To obtain the time-resolved lightcurve of
GX\,17+2, the other sources in the field were subtracted from the {\it
uv}-data before measuring the source flux density using the
\textsc{aips} task \textsc{uvplt}.

\section{Results and discussion}

\subsection{Radio emission correlated with the position in the HID}
\label{radioHID}

\begin{figure}
\begin{center}
\includegraphics[angle=0,scale=0.24]{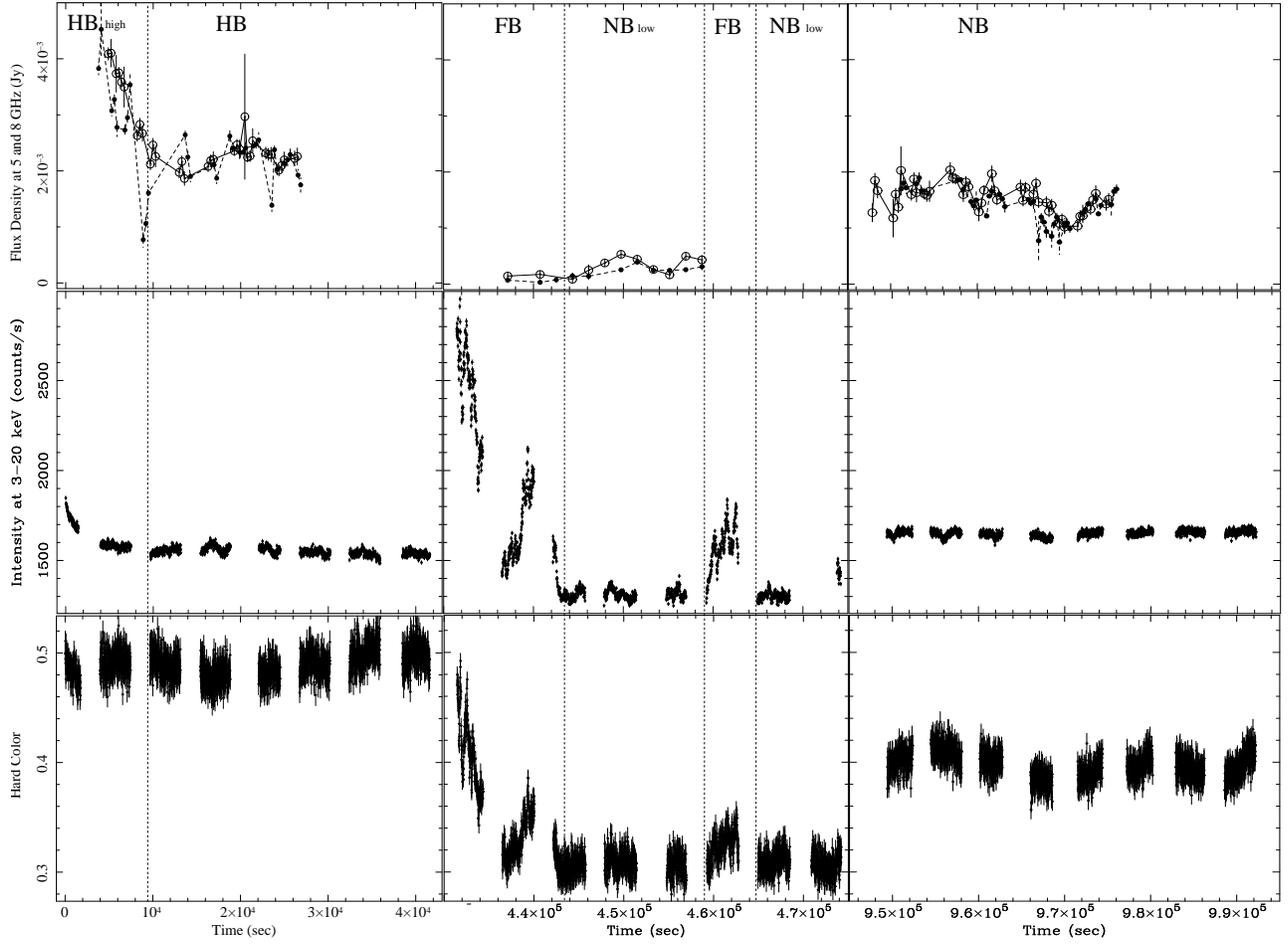}
\caption{Radio and X-ray observations of GX~17+2
  in November 2002. {Top panels:} 5~GHz (open marker, solid line) 8.5~GHz
  (filled marker, dashed line) VLA light curves. {Middle panels:} 3-20 keV PCA/RXTE
  light curves. {Lower panels:} X-ray hardness ratio [(10.5/19.6) keV/(7.1-10.5) keV]. In
  the top panels we also show the X-ray states of the source during
  the observations. }
\label{licu}
\end{center}
\end{figure}

In Fig.~\ref{licu} we show the 5 GHz and 8.5 GHz VLA (top), 3-20 keV
RXTE/PCA (middle) and hard X-ray color (bottom) light curves of GX
17+2.  The simultaneous X-ray and radio light curves show no obvious
correlations between the X-ray count rate and the radio flux
densities. However, during the radio observations, we note an overall
correlation between the mean radio flux density and the mean hard
color (top and bottom panels): the higher the mean radio flux density,
the higher the hard color. More specifically, the radio emission is
strongly related to the position in the HID. In Fig.~\ref{hid}, we
show the HID of GX~17+2: the gray dots are the PCA observations with a
time resolution of 16 seconds and the superimposed open circles
indicate the radio emission strength: the bigger the circle, the more
radio flux density. We clearly see that the radio flux density is
strictly related to the position in the HID, in a way consistent with
previous observations (Penninx et al. 1988) and with the disc-jet
coupling scenario described in Migliari \& Fender (2006): the radio
flux density increases from the FB to the NB and is highest in the
HB. We also note an enhancement of radio flux density in the HB
(i.e. what we called HB$_{high}$), corresponding to an increase in
intensity of the source, with no significant change in hard color.

The overall radio emission is consistent with being optically thick
exept possibly for the observations during the HB$_{high}$, which
correspond to what seems to be the decay of an optically thin radio
flare (with $\alpha=-0.16\pm0.01$, measured by fitting the flux
densities at 1.4, 5, 8.5 and 15~GHz, of the averaged observations for
the range of times $<10^{4}$~sec from the beginning of our
observations; see Fig.~\ref{licu}, top-left panel). The radio decay in
the top-left panel of Fig.~\ref{licu}, {\it might} be associated with
the (preceding) X-ray count rate decay shown in the middle-left
panel. In this case, a possible scenario would be that of a radio
flare associated with the NB-to-HB transition (not observed), of which
we only see the decaying part. However, the X-ray count rate decay can
also be the decay of a long type-I X-ray burst of the same kind
observed for this source in previous RXTE observations (see e.g. the
long `burst 6' reported in Kuulkers et al. 2002). Indeed, the X-ray
decay shown in the PCA light curve seems to be exponential, supporting
the interpretation as a type-I burst.  We tried to investigate this
possibility by fitting the 2-25 keV PCA energy spectrum of the X-ray
decay (the first 1700 seconds of the observation 70023-01-01-01) with
the model described in \S~2.1, without the power-law component, plus
an extra blackbody component to account for the possible thermal
emission from the surface of the NS, but the additional blackbody does
not give a significant statistical improvement to the fit.  The
radio/X-ray flux decay simultaneity would be, in the case of a type-I
burst, only accidental.

\subsection{A  radio jet switches-on $at$ the FB-to-NB state transition}
\label{jetswitch}

\begin{figure}
\begin{center}
\includegraphics[angle=0,scale=0.65]{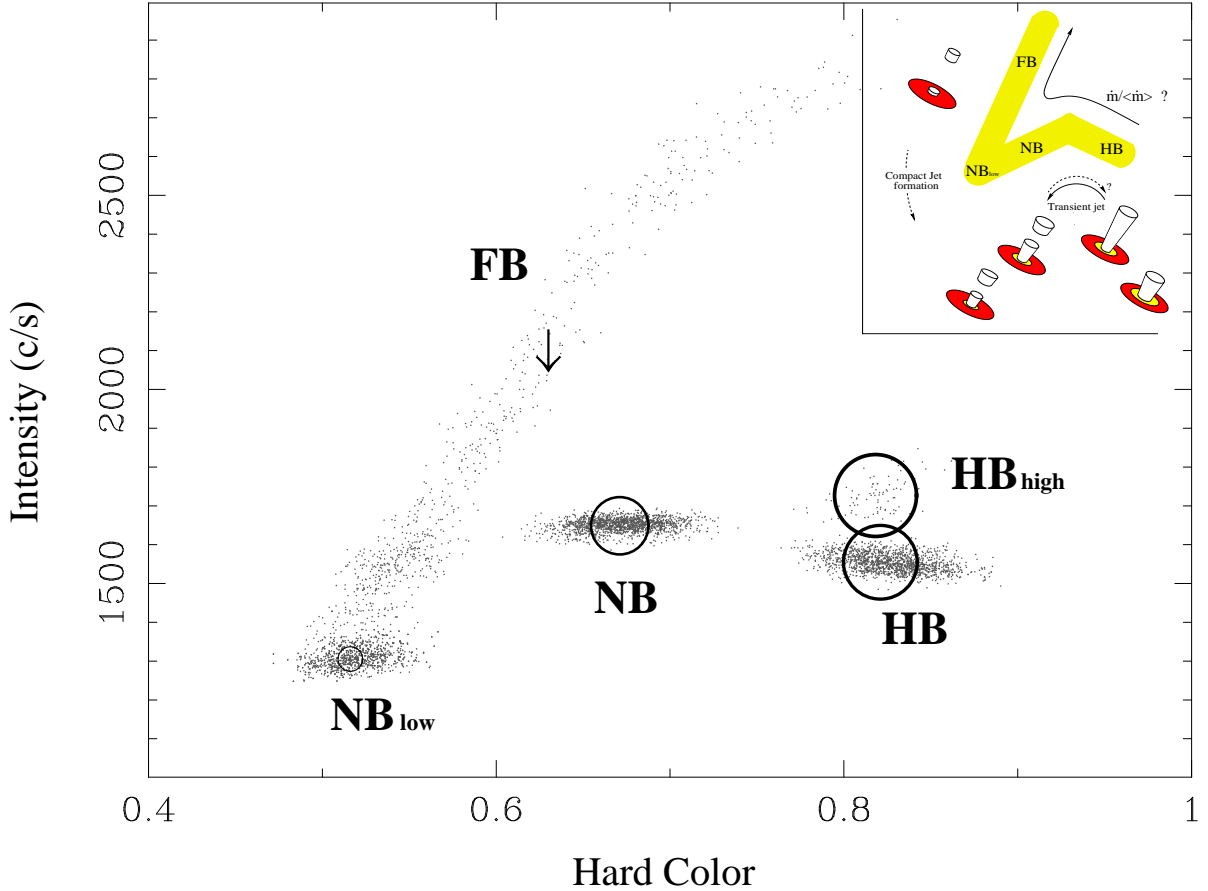}
\caption{Hardness intensity diagram (HID; see \S~2.1) of GX~17+2 (main panel), with a sketch of the
 jet/X-ray state coupling (top-right panel) adapted from Migliari \&
 Fender (2006) to include the formation of the compact jet at the
 FB-to-NB transition (see \S~3.3).  For clarity, the
sketches of the jets in the top-right panel refer only to the cycle
from HB to FB. Indeed, there should be a hysteresis cycle, where, if a
compact jet is reformed during the FB-to-NB, optically thin shocks and
transient jets might not be present. In the main panel, the gray dots
 represent the HID of 16~s of observation. The open circles indicate
 the mean radio flux density of the source observed in the different
 branches: the bigger the circle, the higher the observed radio
 flux density. The arrow in the FB indicates an upper limit on the radio
 flux density. The radio emission is strongest in the HB$_{high}$ and, following
 the HID track, decreases towards the NB$_{low}$, until it is not
 detectable anymore in the FB.}
\label{hid}
\end{center}
\end{figure}

\begin{figure}
\begin{center}
\includegraphics[angle=0,scale=0.35]{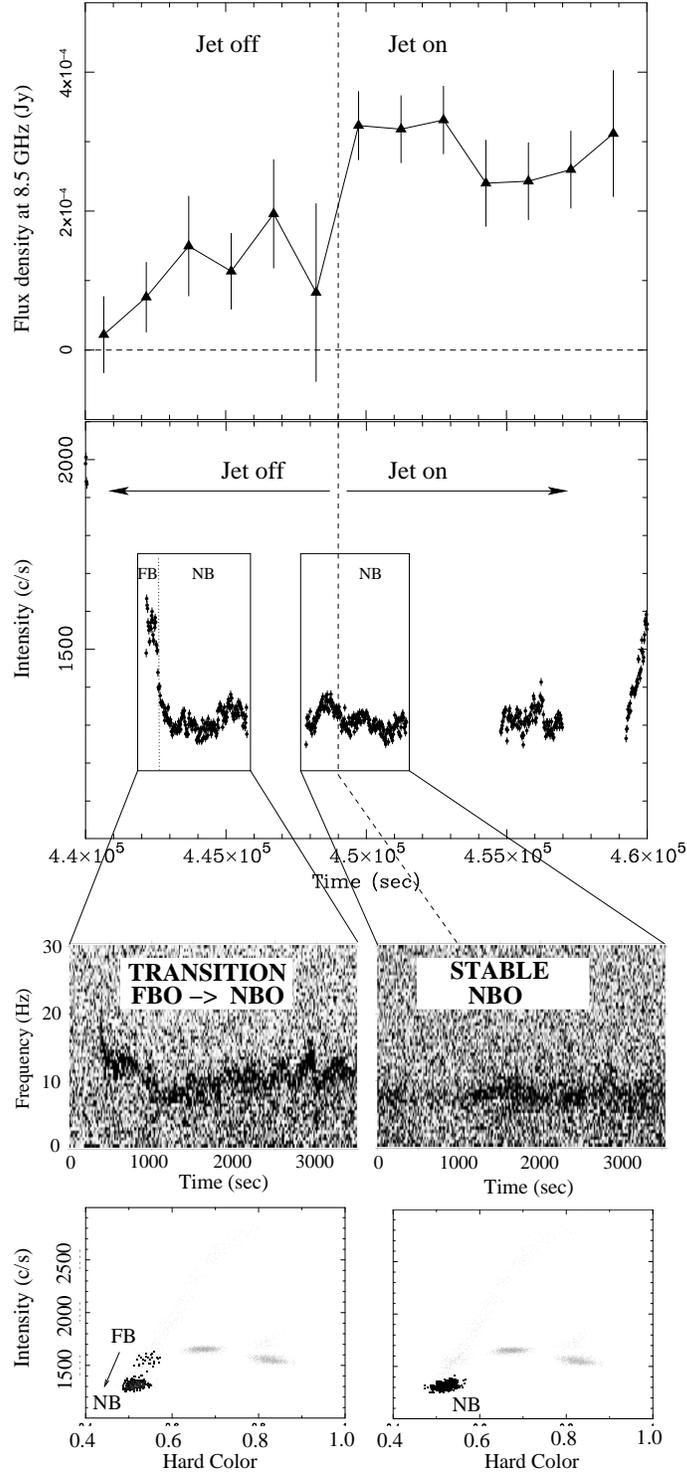}
\caption{From the top: Radio flux density at 8.5~GHz, X-ray light curve, dynamical power density
spectra, and HID of the two RXTE orbits showing a FB-to-NB X-ray
transition and soon after the formation of a compact jet. The
black trace in the dynamical power density spectra show, on the left,
the transition from a FBO to a NBO, and on the right a steady NBO. The
darker points in the lower panels represent the position of the two
RXTE orbits on the HID. See \S~3.2 for details.}
\label{jet-form}
\end{center}
\end{figure}

The HID in Fig.~\ref{hid} shows that during the FB, no radio emission
is detected, while emission, although weak, is observed in the lower
NB. Our observations cover the exact time in which the source transits
from the FB to the lower NB (panels of the middle-column in
Fig.~\ref{licu}).  In Fig.~\ref{jet-form}, we show the radio and X-ray
light curves of GX~17+2 during the transition from FB to NB. The first
panel from the top shows the radio light curve at 8.5 GHz (re-binned
to a higher temporal resolution with respect to that in
Fig.~\ref{licu}). The second panel shows the X-ray PCA light curve.
In the first part of the observation, i.e. before
$4.47\times10^{5}$~sec the source is not significantly detected (with
a nominal flux density of $0.024\pm 0.048$~mJy). However, after
$4.48\times10^{5}$~sec the radio source is detected at 8.5 GHz at a
significance level of $8.8\sigma$ and at 5~GHz at a significance level
of $6.8\sigma$. The radio power-law spectral index is
$\alpha=-0.27\pm0.33$).  Physically, we interpret this as a compact
jet that switches-on soon after the FB-to-NB transition. (Note that,
since the radio spectral index is not well constrained, an optically
thin radio flare cannot be ruled out.)  A sketch of the
(phenomenological) jet/X-ray state coupling model for Z sources
adapted from Migliari \& Fender (2006) to include this result is shown
in the top-right panel of Fig.~\ref{hid}. For clarity, the drawings of
the jet in the top-right panel refer only to the cycle from HB to
FB. Indeed, if a compact jet is reformed during the FB-to-NB
transition, optically thin shocks and transient jets may not be
present in the NB.

In Fig.~\ref{jet-form}, the third panels from the top show the
dynamical power density spectra of the two orbits (in black are
evident the changes in the characteristic frequency of the QPOs), and
the bottom panels show the positions of the observations on the HID as
black markers. The lower FB is usually characterized by the presence
of the so-called flaring branch oscillation (FBO), which has a typical
characteristic frequency above 14~Hz, while the NB power density
spectra usually show the so-called normal branch oscillation (NBO),
which has a typical characteristic frequency below 10~Hz (see Homan et
al. 2002). The first RXTE/PCA orbit shows the actual X-ray state
transition from the FB to the lower NB: the black markers in the HID
are spread in between the FB and the NB, and the dynamical power
density spectrum shows a transition between the FBO and the NBO, with
the QPO frequency oscillating between the typical frequencies of the
two QPOs; the fit of the averaged power density spectrum gives a
characteristic frequency of $\nu_{QPO}=10.7\pm0.1$ (the same kind of
`intermediate' QPO has been already observed in previous observations
of GX~17+2: Homan et al. 2002). The QPO stabilizes into a NBO only in
the second orbit with a characteristic frequency of
$\nu_{NBO}=7.8\pm0.1$~Hz. This stable NBO is simultaneous with a clear
renewed radio activity of the source. As the QPO frequency stabilizes
into a NBO, the compact jet appears to re-establish itself.  The
parallel between the stability of the NBO and the radio emission, and
since in Z-sources the NBO, as well as the radio emission, has been
observed in all the X-ray states with the exeption of the FB, suggests
a relation between the formation of the jet and the X-ray variability
associated to the NBO. In an attempt to draw analogies between the
fast X-ray variability observed in Z sources and in BH XRBs, Casella,
Belloni \& Stella (2005) studied the properties of the low-frequency
QPOs in the two types of systems, and related the FBO to the so-called
`type-A' QPO, the NBO to the `type-B' QPO and the HBO to the `type-C'
QPO. In this framework, the jet formation in BHs might be related to
the type-B QPO, and/or to the transition between the type-A to the
type-B QPO.  An association between the presence of QPOs and radio jet
activity, similar to what we see in GX~17+2, can be found in some BH
XRBs, albeit with a much slower time-scale. A clear example comes from
the multiwavelength studies of the decay of the outburst in H1743-322
(Kalemci et al. 2006). The power density spectrum of the source shows
a low-frequency QPO that appears during the transition from the
thermal/soft state, where the radio jet is undetected, to the hard
X-ray state (see also e.g. Homan \& Belloni 2005). While the source is
in an intermediate state and entering the hard state, the
characteristic frequency of the QPO decreases in time (possibly also
changing `type'; in their work Kalemci et al. did not classify the
QPOs by `type') and, when the source is in its hard state, the radio
jet renews its activity.

In the bright atoll source GX~13+1, Homan et al. (2004) observed a
delay of approximately 40 minutes between the changes in the X-ray
spectral hardness and the following radio flares.  In GX~17+2, we
observe renewed radio activity around $4.5\times10^{5}$~sec
(Fig.~\ref{jet-form}, top). If we associate this radio flux increase
with the preceding FB-to-NB transition, the delay is $\sim2$~hr. This
time-delay is about three times that found in GX~13+1 and about twice
the delays observed in the highly-accreting BH XRB GRS~1915+105
between the beginning of the X-ray hard dips and the following radio
flares (see Klein-Wolt et al. 2002), i.e. taking the hard X-ray dips'
duration as the time that it takes the compact jet to re-form.  If, on
the other hand, the radio activity is associated with the stability of
the characteristic frequency of the NBO, we obtain an upper limit on
the X-ray/radio activity delay of $\lesssim4000$~s, consistent with
what has been observed in the NS GX~13+1 and the BH GRS~1915+105.  For
comparison, more `traditional' BHs like H1743-322 or 4U~1543-47
(Kalemci et al. 2005, 2006), during the decay of the outburst show a
time delay between the end of the thermal/soft state and the detection
of the radio jet of at least a few days.

\subsection{Radio emission and hard X-ray tails}
\label{hardtail}

\begin{figure}
\begin{center}
\includegraphics[angle=0,scale=0.5]{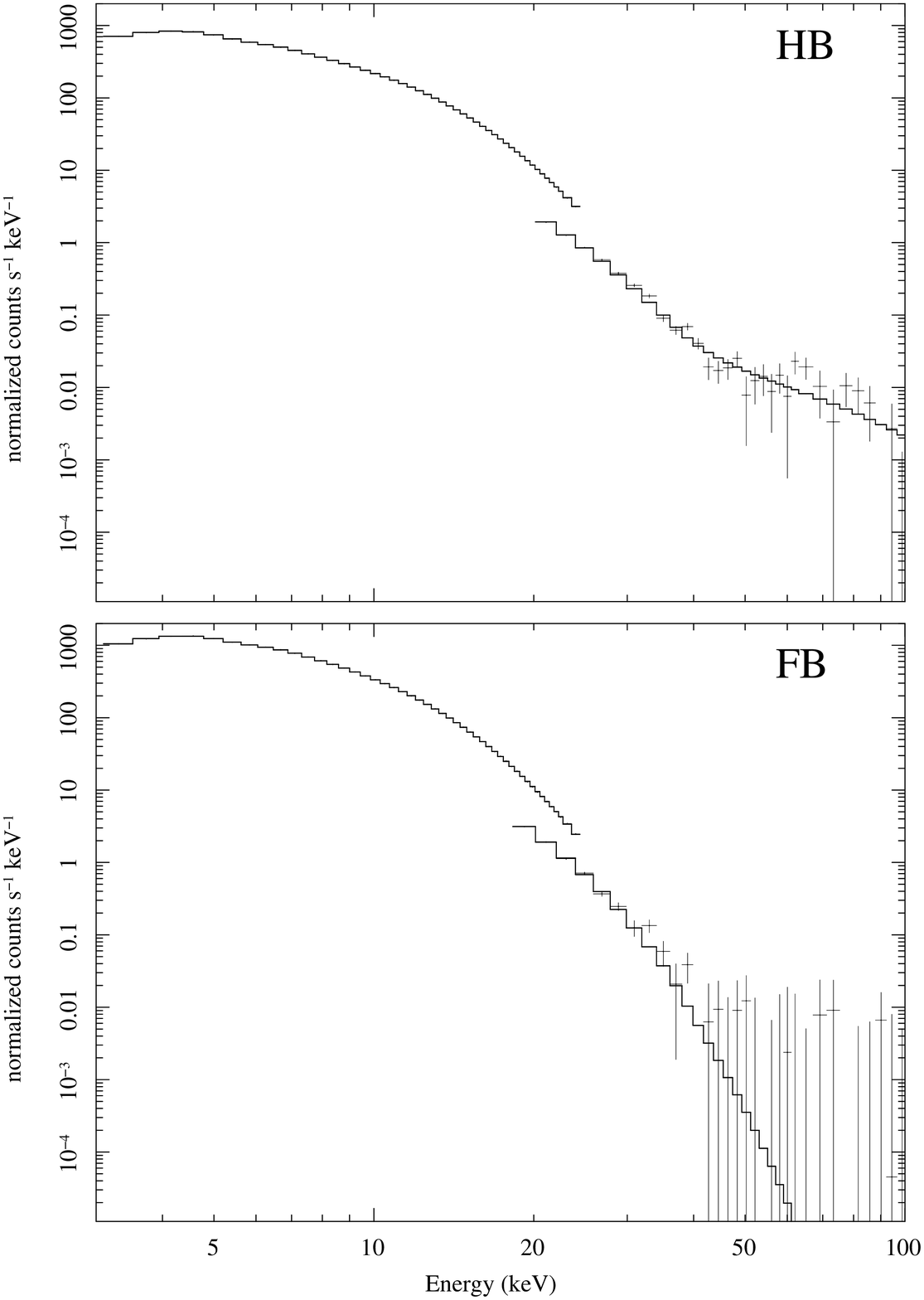}
\caption{X-ray spectra of GX~17+2 in the HB (top panel) and the FB
(lower panel), showing the presence of the hard X-ray tail above
$\sim30$~keV in the HB.}
\label{tail}
\end{center}
\end{figure}

We have analysed the 3-100 keV energy spectra of five observations
along the HID. In Table~2 we show the best-fit parameters of the X-ray
energy spectra using the model described in \S~2.1. A hard X-ray tail
is present in the HB$_{high}$ and HB (an F-test for the addition of a
power-law component gives a chance improvement probability of
$\sim5\times10^{-7}$ in HB$_{high}$ and $\sim3\times10^{-24}$ in HB;
Fig.~\ref{tail}, top panel), and the 2--100~keV flux in the
power-law component is approximately 15\% and 10\%, respectively, of
the total 2-100 keV flux of the source. In the NB an extra power-law
component is also needed in the fit of the energy spectrum (an F-test
gives $\sim9\times10^{-6}$ chance probability), but the power-law flux
decreases to approximately 1\% of the total 2-100 keV flux.  No
additional component is needed to the spectral fits when the source is
in NB$_{low}$ and in the FB (Fig.~\ref{tail}, lower panel). Therefore,
we confirm that in GX~17+2 there is a clear correspondence between the
presence of a hard tail in the X-ray spectra and the position on the
HID.

\begin{figure}
\begin{center}
\includegraphics[angle=-90,scale=0.5]{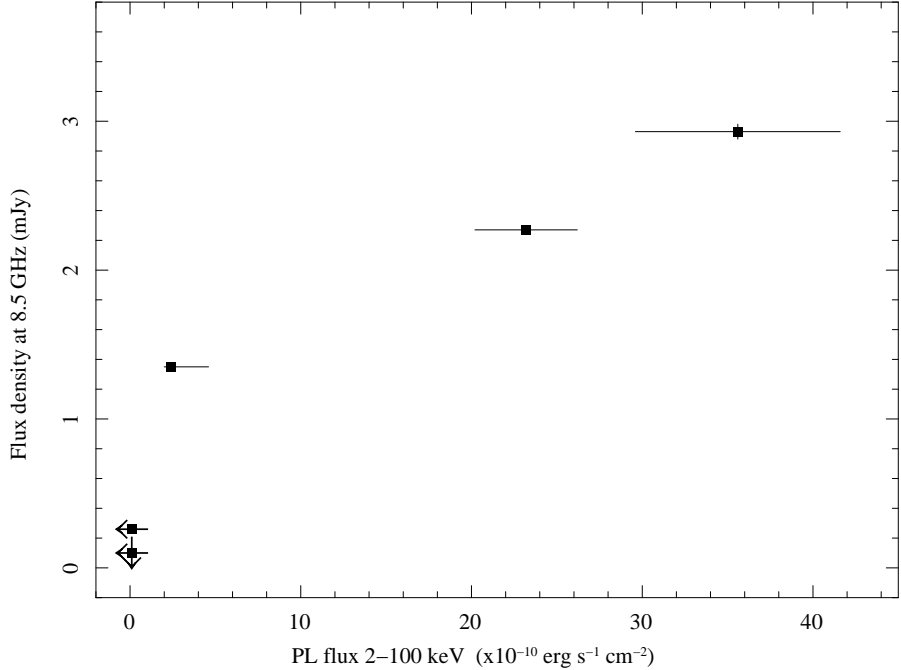}
\caption{X-ray flux in the hard tail power law component as a function
 of the 8.5~GHz radio flux density for observations in five different
 positions of the HID.}
\label{radio-tail}
\end{center}
\end{figure}

The behaviour of the hard X-ray tail in GX~17+2 as a function of the
position in the HID is qualitatively the same as that observed for the
radio emission, suggesting a common physical driver for the production
of the radio and hard X-ray tail emission. In order to quantify this
qualitative radio emission/hard tail correspondence, in
Fig.~\ref{radio-tail} we plot the mean radio flux density at 8.5 GHz
against the 2-100 keV flux of the hard tail power-law component for
the observations in FB, NB$_{low}$, NB, HB, HB$_{high}$.  The three
observations for which the presence of the X-ray power law and the
radio emission is significantly detected show a positive correlation
between the two quantities (with a correlation coefficient of 99\%):
the radio flux density increases as the power-law X-ray flux
increases. The upper limits on the other two observations are
consistent with this trend.
However, given the present statistics, the correlation cannot be firmly
constrained and other observations are needed in order to confirm and
properly quantify the radio/hard X-ray dependence.

In BH systems, it has been suggested that the Comptonising corona gets
ejected during a radio flare (e.g. Vadawale et al. 2003; Fender,
Belloni \& Gallo 2004); hence the transition of the source into the
soft state.  In the NS system GX~17+2, we appear to see a correlation
between the hard X-ray tail and the radio emission, with a strong hard
X-ray tail also present during a radio flare.  However, although
during the HB$_{high}$ observation what we see is likely the decay
portion of a radio flare, thus associated with a preceding `transient'
ejection, the radio spectra of GX~17+2 during the other X-ray states
are consistent with the emission from a `compact' jet. The compact
jet, which can be present also during the HB$_{high}$, is what appears
to be associated with the hard X-ray tail. Similarly, in BH systems
the radio compact jet is observed in the quiescent/hard state, when also
a hard Comptonizing `corona' is present.

\section{Conclusions}

We have analysed simultaneous radio (VLA) and X-ray (RXTE)
observations covering all the three X-ray branches of the Z-type NS
XRB GX~17+2 and found:\\

\begin{itemize}

\item A relation between the radio emission and the position in the
HID: the radio flux density is strongest in the HB, decreases in the NB,
and is weakest in the FB (Fig.~2).
\item A relation between the presence of a hard X-ray tail and the
position on the HID: the hard power-law tail is observed in the HB, in
the NB, and is not detected in the lower NB and in the FB (Fig.~4 and
Table~2).
\item A link between X-ray state transitions and the jet activity: a
 jet, likely a compact jet, forms soon after the FB-to-NB transition,
 with a time delay of less than 2~hr (Fig.~3).
\item The radio emission of the jet (re-)formed after the FB-to-NB state transition stabilizes
when the NBO characteristic frequency in the X-ray power spectrum also
stabilizes (Fig.~3). This finding, together with the fact that in
Z-sources the NBO, as well as the radio emission, has been observed in
all the X-ray states with the exception of the FB, suggest a relation
between the formation of the jet and the X-ray variability associated
to the NBO. A similar behaviour, albeit with a longer time-scale, may
be found in BH systems, where the decrease of the characteristic
frequency and change of `type' of the low-frequency QPO is followed by
a renewed activity of the jet (see \S~\ref{jetswitch}).
\item An indication for a quantitative relation between radio
emission and the hard tail in the X-ray spectrum: there is a positive
correlation between the hard tail power-law X-ray flux and the radio
flux density (Fig.~5). If further confirmed with a larger sample and
improved statistics, especially in the hard X-ray tail flux
measurement, this relation would point to a common mechanism for the
production of the jet and the hard X-ray tails in the system.

\end{itemize}

\acknowledgments
	
SM would like to thank Tommy Thompson for useful discussions.  The
National Radio Astronomy Observatory is a facility of the National
Science Foundation operated under cooperative agreement by Associated
Universities, Inc.

\begin{table}
\caption{Log of the RXTE observations of GX~17+2 simultaneous with the VLA.}
\begin{tabular}{l|l|l|l}
\tableline
\tableline  
Obs ID & Start time & End time & X-ray states   \\
\tableline
70023-01-01-01 & 2002-11-04UT18:19:28 & 2002-11-04UT22:03:28 & HB$_{high}$/HB\\
70023-01-01-03 &2002-11-04UT22:37:36 & 2002-11-04UT23:37:20 & HB\\
70023-01-01-00 &2002-11-05UT00:28:32 & 2002-11-05UT06:02:08 & HB \\
70023-01-02-00 &2002-11-09UT18:11:44 &2002-11-09UT23:46:40 & FB/NB$_{low}$\\
70023-01-02-01 & 2002-11-10UT00:40:00& 2002-11-10UT06:07:12& FB/NB$_{low}$\\
70023-01-03-00 & 2002-11-15UT18:02:40 & 2002-11-15UT21:44:32 & NB\\
70023-01-03-01 & 2002-11-15UT22:40:00 & 2002-11-16UT05:57:20 & NB\\

\tableline

\end{tabular}
\end{table}

\begin{table}
\caption{Best-fit parameters of the observations corresponding to five
different positions on the HID: for the HB we analysed the observation
70073-01-01-00, for HB$_{high}$ the first two orbits of
70073-01-01-01, for NB we averaged 70073-01-03-00 and 70073-01-03-01,
for FB and NB$_{low}$ we analysed the first and the fourth orbit of
70073-01-02-00, respectively. Errors are at the 68\% confidence level
for a single parameter. We fitted the 3-100 keV spectra using a model
consisting of a blackbody (kT$_{BB}$ is the blackbody temperature,
N$_{BB}$ is the normalization in units of L$_{39}$/D$_{10}^{2}$ where
L$_{39}$ is the luminosity of the source in units of $10^{39}$ erg/s
and D$_{10}$ is the distance to the source in units of 10~kpc), a
Gaussian emission line (E$_{Fe}$ is the line energy in keV, EqW$_{Fe}$
is the line equivalent width in keV, $\sigma_{Fe}$ is the line width
in keV, N$_{Fe}$ is the total photons cm$^{-2}$ s$^{-1}$ in the line),
CompTT (kT$_{W}$ is the input soft photon Wien temperature in keV,
kT$_{e}$ is the plasma temperature in keV, $\tau$ is the plasma
optical depth), an absorption edge (E$_{edge}$ is the threshold energy
in keV, Max~$\tau$ is the maximum absorption factor at threshold), a
power law ($\Gamma_{PL}$ photon index, N$_{PL}$ is the normalization
in photons keV$^{-1}$ cm$^{-2}$ s$^{-1}$ at 1 keV) when needed, all
corrected for photoelectric absorptions (N$_{\rm H}$). F-test is the
probability of chance improvement for the addition of a power law
component. Flux$_{PL}$ is the 2-100~keV flux of the additional power-law
component in units of $10^{-10}$~erg cm$^{-2}$ s$^{-1}$.}

\begin{tabular}{l|l|l|l|l|l}
\tableline
\tableline  
Parameter         & HB$_{high}$ & HB & NB & NB$_{low}$ & FB \\
\tableline

N$_{\rm H}$ ($\times 10^{22}$ cm$^{-2}$)&   2 (fixed)      &  2 (fixed)  &  2 (fixed)  & 2 (fixed)  & 2 (fixed)\\

kT$_{BB}$ (keV)                    &  $0.55\pm0.07$          & $0.46^{+0.03}_{-0.11}$ & $0.57^{+0.01}_{-0.06}$ &$0.53^{+0.09}_{-0.07}$  & $0.80^{+0.07}_{-0.13}$\\

N$_{BB}$  $(\times 10^{-2})$       &  $0.87^{+0.04}_{-0.02}$ & $2.37^{0.23}_{0.17}$          & $3.93^{+0.18}_{-0.06}$ &$4.74^{+0.38}_{-0.48}$  & $9.08^{+0.25}_{-0.15}$\\

kT$_{W}$                           & $0.82^{+0.07}_{-0.01}$  &$0.78\pm0.01$           & $0.90\pm0.01$          &$0.96\pm0.01$           & $1.35\pm0.01$\\

kT$_{e}$                           & $3.08^{+0.01}_{-0.02}$  & $3.21\pm0.01$          & $2.97\pm0.01$          &$2.93\pm0.02$           & $2.86\pm0.01$\\

$\tau$                             &  $6.29\pm0.03$          & $6.06^{+0.01}_{-0.03}$ & $5.24\pm0.01$          &$4.16^{+0.09}_{-0.06}$  & $5.57^{+0.03}_{-0.04}$\\

N$_{Comp}$                         &  $1.06^{+0.09}_{-0.07}$ & $0.97\pm0.01$          & $1.24\pm0.01$          &$1.06\pm0.01$           & $1.34\pm0.02$\\

E$_{Fe}$                           &  $6.58^{+0.07}_{-0.06}$ &  $6.52\pm0.06$         & $6.52\pm0.06$          &$6.60\pm0.04$           & $6.51\pm0.05$\\

EqW$_{Fe}$                 & $0.08\pm0.01$ & $0.07\pm0.01$    &   $0.07\pm0.01$  &  $0.13\pm0.01$  & $0.09\pm0.01$\\

$\sigma_{Fe}$                      & $0.41^{+0.08}_{-0.17}$  & $0.28^{+0.13}_{-0.10}$ & $0.33^{+0.11}_{-0.09}$ &$0.46^{+0.08}_{-0.10}$  & $0.36^{+0.14}_{-0.11}$\\

N$_{Fe}$   $(\times 10^{-2})$      & $1.28^{+0.14}_{-0.17}$  & $1.03^{+0.12}_{-0.09}$ & $1.08^{+0.16}_{-0.06}$ &$1.65^{+0.18}_{-0.21}$  & $2.22^{+0.35}_{-0.29}$\\

E$_{edge}$                         & $9.2\pm0.2$             & $9.06^{+0.01}_{-0.01}$ & $9.18^{+0.16}_{-0.15}$ &$9.42^{+0.14}_{0.15}$   & $9.04^{+0.16}_{-0.17}$\\

Max~$\tau$    $(\times 10^{-2})$    & $3.05^{+0.49}_{-1.34}$ & $3.89^{+0.49}_{-0.51}$& $3.27^{+0.37}_{-0.36}$ &$4.16^{+0.69}_{-0.34}$  & $4.06^{+0.90}_{-0.73}$\\

$\Gamma_{PL}$                      & $2.88^{+0.04}_{-0.02}$   & $2.77^{+0.03}_{-0.01}$& $2.79^{+0.21}_{-0.03}$ & $\dots$                   & $\dots$\\

N$_{PL}$                           & $3.73^{+0.47}_{-0.74}$   & $1.99^{+0.27}_{-0.31}$  & $0.21^{+0.20}_{-0.04}$ & $\dots$                   & $\dots$\\

Flux$_{PL}$ &$36\pm6$ & $23\pm3$ & $2.4^{+2.2}_{-0.4}$& $<0.1$& $<0.1$\\

$\chi^{2}_{red}$ (d.o.f.)          &  1.17 (95)              & 1.22 (95)              & 1.33 (95)              & 0.97 (99)              & 0.86 (99)\\

F-test (+PL)                       &  $5.1\times 10^{-7}$    & $3.0\times10^{-24}$    & $8.9\times10^{-6}$     & $\dots$                   & $\dots$\\

\tableline

\end{tabular}
\end{table}

\end{document}